\documentclass[prd,showpacs,superscriptaddress,nofootinbib,floatfix,showkeys,10pt]{revtex4-2}
\usepackage{bm}
\usepackage{times}
\usepackage{braket}
\usepackage{amsfonts,amssymb,stmaryrd,latexsym,amsmath}
\usepackage[usenames,dvipsnames]{color}
\usepackage{epsfig}
\usepackage{slashed}
\usepackage{hyperref}
\usepackage{subfigure}
\usepackage{graphics}
\usepackage{slashed}
\usepackage{orcidlink}
\usepackage{multirow}
\usepackage{dashrule}
\allowdisplaybreaks[1]

\allowdisplaybreaks[1]

\definecolor{nicered}{rgb}{0.7,0.1,0.1}
\definecolor{nicegreen}{rgb}{0.1,0.5,0.1}
\definecolor{emph}{rgb}{1,0,0}
\definecolor{doub}{rgb}{0.7,0.2,1.0}
\definecolor{navyblue}{RGB}{0, 110, 184}

\hypersetup{colorlinks,citecolor=nicegreen,linkcolor=nicered,urlcolor=navyblue}

\newcommand{\clabel}[2][]{#2}

\begin{document}

	
\title{Electromagnetic polarizabilities of the spin-$\frac{1}{2}$ singly heavy baryons in heavy baryon chiral perturbation theory} 

\author{Yan-Ke Chen\,\orcidlink{0000-0002-9984-163X}}
\affiliation{School of Physics, Peking University, Beijing 100871, China}

\author{Liang-Zhen Wen\,\orcidlink{0009-0006-8266-5840}}\email{wenlzh\_hep-th@stu.pku.edu.cn}
\affiliation{School of Physics, Peking University, Beijing 100871, China}

\author{Lu Meng\,\orcidlink{0000-0001-9791-7138}}\email{lu.meng@rub.de}
\affiliation{Institut f\"ur Theoretische Physik II, Ruhr-Universit\"at Bochum,  D-44780 Bochum, Germany }

\author{Shi-Lin Zhu\,\orcidlink{0000-0002-4055-6906}}\email{zhusl@pku.edu.cn}
\affiliation{School of Physics and Center of High Energy Physics,
Peking University, Beijing 100871, China}

\begin{abstract}

We calculate the electromagnetic polarizabilities of the spin-$\frac{1}{2}$ singly heavy baryons in the heavy baryon chiral perturbation theory up to $\mathcal{O}(p^3)$. 
We estimate the low-energy constants using the magnetic moments of singly charmed baryons from lattice QCD simulations and the experimental decay widths of $\Sigma_c$ and $\Sigma_c^*$. Our results indicate that the long-range chiral corrections make significant contributions to the polarizabilities. Additionally, the magnetic dipole transitions $\mathcal{B}_6^* \to \mathcal{B}_6 +\gamma $ also provide large contribution to the magnetic polarizabilities.
\end{abstract}
 
\maketitle


\section{Introduction}

The electromagnetic properties of heavy baryons at low energies are highly sensitive to their internal structures and the chiral dynamics of light quarks. These properties provide valuable insights into the mysteries of nonperturbative QCD and confinement effects. In particular, the electromagnetic properties such as radiative decays and magnetic moments have been extensively studied using various theoretical approaches, including quark models~\cite{Barik:1983ics,Ivanov:1996fj,Ivanov:1999bk,Tawfiq:1999cf,Julia-Diaz:2004yqv,Kumar:2005ei,Faessler:2006ft,Sharma:2010vv,Majethiya:2011ry,Wang:2017kfr,Hazra:2021lpa}, bag models~\cite{Bose:1980vy,Simonis:2018rld,Bernotas:2012nz,Bernotas:2013eia,Zhang:2021yul}, QCD sum rules~\cite{Zhu:1997as,Zhu:1998ih,Aliev:2008ay,Aliev:2008sk,Wang:2010xfj,Aliev:2011bm,Agamaliev:2016fou,Ozdem:2024brk}, the Skyrme model~\cite{Oh:1991ws,Oh:1995eu}, hypercentral models~\cite{Patel:2007gx}, pion mean-field approaches~\cite{Yang:2018uoj,Yang:2019tst,Kim:2021xpp}, bound-state approaches~\cite{Scholl:2003ip}, chiral perturbation theory ($\chi$PT)~\cite{Cheng:1992xi,Cho:1994vg,Savage:1994wa,Banuls:1999br,Tiburzi:2004mv,Jiang:2015xqa,Li:2017cfz,Meng:2017dni,Wang:2018gpl,Meng:2018gan,Wang:2018cre,Shi:2018rhk,Wang:2019mhm,Shi:2021kmm}, and lattice QCD simulations\cite{Can:2013tna,Can:2015exa,Bahtiyar:2015sga,Bahtiyar:2016dom,Bahtiyar:2018vub,Can:2021ehb}. Recent reviews~\cite{Can:2021ehb,Meng:2022ozq} provide a broader overview of theoretical advancements in this area.

Experimentally, the short lifetimes of heavy baryons pose significant challenges to direct measurements of their electromagnetic dipole moments. However, advancements in experimental techniques are making such measurements increasingly feasible. For example, the LHC has proposed a method to measure the electromagnetic dipole moments of short-lived charmed baryons by utilizing spin precession of channeled particles in bent crystals~\cite{Aiola:2020yam}. 

 Despite these advancements, there has been relatively little theoretical research on the electromagnetic polarizabilities of heavy baryons, even though these are also critical electromagnetic properties.  With continued advancements in experimental techniques, it may become possible to measure the polarizabilities of heavy baryons. Therefore, conducting theoretical calculations of heavy baryon polarizabilities in advance is both timely and significant. 

 The electric ($\alpha_E$) and magnetic ($\beta_M$) polarizabilities are fundamental constants that describe the response of a system to the application of an external quasistatic electric or magnetic field~\cite{Holstein:2013kia}. The simplest case is the scalar polarizabilities in classical physics, where the electric (magnetic) polarizability of a system is simply the constant of proportionality between the applied static and uniform electric field $\bm{E}$ (magnetic field $\bm{H}$), and the induced electric dipole moment $\bm{d}$ (magnetic dipole moment $\bm{m}$):
 \begin{equation}
 	\bm{d}=4\pi\alpha_E \bm{E},\quad \bm{m}=4\pi\beta_M \bm{H},
 \end{equation}
which leads to a potential energy:
\begin{equation}
	U_E=-\frac{1}{2} 4 \pi \alpha_E \bm{E}^2,\quad U_H=-\frac{1}{2} 4 \pi \beta_M \bm{H}^2.
\end{equation}
This concept can be extended to hadronic systems. The electromagnetic polarizabilities of a particle can be defined through the second-order effective Hamiltonian:
\begin{equation}\label{eq:second_effective_Hamiltonian}
	H^{(2)}=-\frac{1}{2} 4 \pi \alpha_E \bm{E}^2-\frac{1}{2} 4 \pi \beta_M \bm{H}^2.
\end{equation}
For a zero-spin particle with charge $Ze$, the effective Hamiltonian in Eq.~\eqref{eq:second_effective_Hamiltonian} corresponds to a Compton scattering amplitude~\cite{Bernabeu:1976jq,Llanta:1979kj}:
\begin{equation}
	T^{(2)}=\boldsymbol{\epsilon} \cdot \boldsymbol{\epsilon}^{\prime}\left(\frac{-Z^2 e^2}{M}+\omega \omega^{\prime} 4\pi \alpha_E\right)+(\boldsymbol{\epsilon} \times \bm{k}) \cdot\left(\bm{\epsilon}^{\prime} \times \bm{k}^{\prime}\right) \omega \omega^{\prime} 4\pi \beta_M+\mathcal{O}\left(\omega^4\right),
\end{equation}
where $\bm{k}^{(\prime)}$, $\omega^{(\prime)}$, and $\bm{\epsilon}^{(\prime)}$ represent the momentum, energy, and polarization of the initial (final) photon, respectively. Therefore, the electromagnetic polarizabilities of a particle can be determined by studying its Compton scattering amplitude. A generalized expression for the electromagnetic polarizabilities of a particle $|i\rangle$ can be derived from second-order perturbation theory~\cite{Ericson:1973dtc}:
\begin{equation}~\label{eq:generalized_alpha_beta}
	\alpha_E=2 \alpha_{\mathrm{em}}\sum_{n \neq i} \frac{\left|\left\langle n\left| D_z\right| i\right\rangle\right|^2}{E_n-E_i},\quad \beta_M=2 \alpha_{\mathrm{em}}\sum_{n \neq i} \frac{\left|\left\langle n\left| M_z\right| i\right\rangle\right|^ 2}{E_n-E_i},
\end{equation}
where we retain only the leading terms coming from second-order perturbation theory in the long-wavelength limit. $D_z$ and $M_z$ are the $z$-components of the electric and magnetic dipole operators, respectively. From Eq.~\eqref{eq:generalized_alpha_beta}, we can observe that the electric (magnetic) polarizability of a particle is closely related to its electric dipole transition $\left\langle n\left| D_z\right| i\right\rangle$ (magnetic dipole transition $\left\langle n\left| M_z\right| i\right\rangle$). Furthermore, near-degenerate energy levels can result in significantly large electromagnetic polarizabilities, which diverge in the case of accidental degeneracies.

From a theoretical perspective, chiral dynamics are expected to contribute significantly to the electromagnetic polarizabilities of matter fields. In the chiral limit, pions are massless, and a single particle of the matter field can become degenerate with states containing additional pions. As a result, electromagnetic polarizabilities would diverge in the chiral limit. With chiral symmetry breaking, one can still expect photons to interact strongly with the pion cloud at low energies, leading to substantial contributions from long-range chiral corrections to the electromagnetic polarizabilities. Thus, electromagnetic polarizabilities provide an excellent platform to investigate chiral dynamics. In this context, chiral perturbation theory serves as a powerful tool for calculating electromagnetic polarizabilities.

In order to calculate the electromagnetic polarizabilities of heavy baryons, we can draw on the successful experiences in the nucleon system. In Refs.~\cite{Schumacher:2005an,Holstein:2013kia,Hagelstein:2020vog}, the authors reviewed the theoretical and experimental progress in nucleon electromagnetic polarizabilities (see Refs.~\cite{Fonvieille:2019eyf,Sparveris:2024kjz} for the review about proton electromagnetic generalized polarizabilities). Our primary focus here is on studies with heavy baryon chiral perturbation theory (HB$\chi$PT). HB$\chi$PT is a model-independent method for studying baryon properties at the low-energy regime. It addresses the issue in $\chi$PT where the nonvanishing baryon mass in the chiral limit disrupts power counting used in the pure meson sector~\cite{Weinberg:1978kz,Gasser:1983yg,Gasser:1984gg,Jenkins:1990jv,Bernard:1992qa}. For the nucleon polarizabilities, calculations based on HB$\chi$PT at $\mathcal{O}(p^3)$ and $\mathcal{O}(p^4)$ have been performed by Bernard, Kaiser, Schmidt, and Mei\ss ner (BKSM)~\cite{Bernard:1991rq,Bernard:1991ru,Bernard:1993bg,Bernard:1993ry}. Their theoretical results qualitatively capture the features of the polarizabilities and agree with experimental data~\cite{ParticleDataGroup:2020ssz}. The BKSM calculations show that the nucleon polarizabilities primarily come from the contributions of chiral corrections ($N\pi$ loops). One can find more investigations on nucleon electromagnetic polarizabilities in HB$\chi$PT~\cite{Butler:1992ci,Babusci:1996jr,Beane:2004ra,Choudhury:2007qiz} and in covariant baryon chiral perturbation theory~\cite{Lensky:2014efa,Lensky:2015awa,Thurmann:2020mog}. Recent lattice QCD simulations also reveal substantial contributions from the $N\pi$ states to electromagnetic polarizabilities~\cite{Wang:2023omf}. Only when the contributions of the $N\pi$ states are included do the lattice QCD results agree with experimental values, otherwise, the results are significantly smaller. This highlights the importance of the chiral dynamics in electromagnetic polarizabilities. These studies also demonstrate that HB$\chi$PT is a highly effective framework for calculating nucleon electromagnetic polarizabilities. Building on this success, we aim to extend HB$\chi$PT calculations to hadrons containing heavy quarks.

In this work, we calculate the electromagnetic polarizabilities of singly heavy baryons in HB$\chi$PT up to $\mathcal{O}\left(p^3\right)$. We estimate the low-energy constants (LECs) in a similar manner to calculating magnetic moments in Refs.~\cite{Meng:2018gan,Wang:2018gpl,Wang:2018cre}, specifically, using the experimental decay widths of $\Sigma_c$ and $\Sigma_c^*$ as well as the magnetic moments of heavy charmed baryons from lattice QCD simulations.

The paper is arranged as follows: in Sec.~\ref{sec:theoretical_framework}, we introduce our theoretical framework. We discuss the spin-averaged forward Compton tensor in Sec.~\ref{subsec:spin-averaged_forward_Compton_ tensor} and construct the effective Lagrangians in Sec.~\ref{subsec:effective_Lagrangian}. In Sec.~\ref{sec:ANALYTICAL_EXPRESSIONS}, we calculate the analytical expressions of the electromagnetic polarizabilities up to $\mathcal{O}(p^3)$. We give the numerical results in Sec.~\ref{sec:NUMERICAL_RESULTS} and a brief summary in Sec.~\ref{sec:DISCUSSION_AND_CONCULSION}.

\section{theoretical framework}\label{sec:theoretical_framework}

The singly heavy baryon consists of one heavy quark and two light quarks. In SU(3) flavor symmetry, the two light quarks in a singly heavy baryon form either an antisymmetric flavor antitriplet $\bar{3}_f$ or a symmetric flavor sextet $6_f$. With the constraint of Fermi-Dirac statistics, the spin of the $\bar{3}_f$ and the $6_f$ diquarks are $0$ and $1$, respectively. Consequently, the total spin of the antitriplet baryon is $S_{\bar{3}}=\frac{1}{2}$, and the total spin of the sextet baryon can be either $S_{6}=\frac{1}{2}$ or $S_{6}=\frac{3}{2}$. We use the notations $\psi_{\bar{3}},~\psi_{6}$ and $\psi^*_{6}$ to denote the antitriplet, spin-$\frac{1}{2}$ sextet, and spin-$\frac{3}{2}$ sextet baryons, respectively. These baryon fields are represented as~\cite{Yan:1992gz}:

\begin{equation}\label{eq:baryon_multi}
\psi_{\bar{3},c}=\left(\begin{matrix}
0 & \Lambda_c^{+} & \Xi_c^{+} \\
-\Lambda_c^{+} & 0 & \Xi_c^0 \\
-\Xi_c^{+} & -\Xi_c^0 & 0
\end{matrix}\right), \quad \psi_{6,c}=\left(\begin{matrix}
\Sigma_c^{++} & \frac{\Sigma_c^{+}}{\sqrt{2}} & \frac{\Xi_c^{\prime+}}{\sqrt{2}} \\
\frac{\Sigma_c^{+}}{\sqrt{2}} & \Sigma_c^0 & \frac{\Xi_c^{\prime 0}}{\sqrt{2}} \\
\frac{\Xi_c^{\prime+}}{\sqrt{2}} & \frac{\Xi_c^{\prime 0}}{\sqrt{2}} & \Omega_c^0
\end{matrix}\right), \quad \psi_{6,c}^{* \mu}=\left(\begin{matrix}
\Sigma_c^{*++} & \frac{\Sigma_c^{*+}}{\sqrt{2}} & \frac{\Xi_c^{*+}}{\sqrt{2}} \\
\frac{\Sigma_c^{*+}}{\sqrt{2}} & \Sigma_c^{* 0} & \frac{\Xi_c^{* 0}}{\sqrt{2}} \\
\frac{\Xi_c^{*+}}{\sqrt{2}} & \frac{\Xi_c^{* 0}}{\sqrt{2}} & \Omega_c^{* 0}
\end{matrix}\right).
\end{equation}
\begin{equation}\label{eq:baryon_multi_2}
\psi_{\bar{3},b}=\left(\begin{matrix}
0 & \Lambda_b^{0} & \Xi_b^{0} \\
-\Lambda_b^{0} & 0 & \Xi_b^- \\
-\Xi_b^{0} & -\Xi_b^- & 0
\end{matrix}\right), \quad \psi_{6,b}=\left(\begin{matrix}
\Sigma_b^{+} & \frac{\Sigma_b^{0}}{\sqrt{2}} & \frac{\Xi_b^{\prime 0}}{\sqrt{2}} \\
\frac{\Sigma_b^{0}}{\sqrt{2}} & \Sigma_b^- & \frac{\Xi_b^{\prime-}}{\sqrt{2}} \\
\frac{\Xi_b^{\prime 0}}{\sqrt{2}} & \frac{\Xi_b^{\prime-}}{\sqrt{2}} & \Omega_b^-
\end{matrix}\right), \quad \psi_{6,b}^{* \mu}=\left(\begin{matrix}
\Sigma_b^{*+} & \frac{\Sigma_b^{* 0}}{\sqrt{2}} & \frac{\Xi_b^{* 0}}{\sqrt{2}} \\
\frac{\Sigma_b^{* 0}}{\sqrt{2}} & \Sigma_b^{* -} & \frac{\Xi_b^{* -}}{\sqrt{2}} \\
\frac{\Xi_b^{* 0}}{\sqrt{2}} & \frac{\Xi_b^{* -}}{\sqrt{2}} & \Omega_b^{* -}
\end{matrix}\right).
\end{equation}

In the HB$\chi$PT scheme, we decompose the heavy baryon fields into the ``heavy'' and ``light'' components as follows
\begin{equation}\label{eq:heavy_baryon_field}
	\mathcal{B}_n(x)=e^{i M_{\mathcal{B}_n} v \cdot x} \frac{1+\slashed v}{2} \psi_n, \quad \mathcal{H}_n(x)=e^{i M_{\mathcal{B}_n} v \cdot x} \frac{1-\slashed v}{2} \psi_n,
\end{equation}
where $\psi_n$ denotes the heavy baryon field $\psi_3, \psi_6$ or $\psi_6^{* \mu}$, and $\mathcal{B}_n\left(\mathcal{H}_n\right)$ is the light (heavy) component of the corresponding heavy baryon field. $M_{\mathcal{B}_n}$ is the baryon mass, and $v^\mu=(1, \bm{0})$ is the static velocity. The heavy field $\mathcal{H}_n(x)$ is then integrated out in the Lagrangians.

In the following, we discuss the electromagnetic polarizabilities of singly charmed baryons in detail. The calculation process for the polarizabilities of singly bottom baryons essentially follows the same steps as those for singly charmed baryons. We present only the final results for singly bottom baryons, which are provided in Appendix~\ref{appendix:b_baryon_polarizabilities}.

\subsection{Spin-averaged forward Compton tensor}\label{subsec:spin-averaged_forward_Compton_ tensor}
To calculate the electromagnetic polarizabilities of heavy baryons, one has to analyze the spin-averaged forward Compton scattering tensor $\Theta_{\mu \nu}$~\cite{Bernard:1991ru,Bernard:1991rq}:
\begin{equation}
	\Theta_{\mu \nu}=\frac{e^2}{4} \operatorname{Tr}\left[(\slashed p+m) T_{\mu \nu}(p, k)\right],
\end{equation}
where $k$ denotes the photon momentum and $\omega=v \cdot k$. $T_{\mu \nu}(p, k)$ is the Fourier-transformed matrix element of two time-ordered electromagnetic currents,
\begin{equation}
T_{\mu \nu}(p, k)=\int d^4 x e^{i k \cdot x}\left\langle \psi(p)\left|T\left[J_\mu^{\mathrm{e m}}(x) J_\nu^{\mathrm{e m}}(x)\right]\right| \psi(p)\right\rangle.
\end{equation}
In the heavy baryon formalism, $\Theta_{\mu \nu}$ can be expressed as~\cite{Bernard:1993bg,Bernard:1993ry}:
\begin{equation}
\begin{aligned}
\Theta_{\mu \nu} & =\frac{e^2}{4} \operatorname{Tr}\left[(1+\slashed v) T_{\mu \nu}(v, k)\right] \\
& = U(\omega) g_{\mu \nu}+V(\omega) k_\mu k_\nu+W(\omega)\left(k_\mu v_\nu+v_\mu k_\nu\right)+X(\omega) v_\mu v_\nu,
\end{aligned}
\end{equation}
We adopt the ``Coulomb gauge" in this paper, where $\epsilon \cdot v = 0$ for the photon polarization vector $\epsilon$. The auxiliary function $\epsilon^{\prime \mu} \Theta_{\mu \nu} \epsilon^\nu$ then reads:
\begin{equation}
	\epsilon^{\prime \mu} \Theta_{\mu \nu} \epsilon^\nu=\left(\epsilon^{\prime} \cdot \epsilon\right) U(\omega)+\left(\epsilon^{\prime} \cdot k \epsilon \cdot k \right)V(\omega).
\end{equation}
The spin-averaged forward Compton amplitude is correlated with two form factors, $U(\omega)$ and $V(\omega)$.  The form factor $V(\omega)$ can be eliminated for real photons, but it provides information about the magnetic polarizability $\beta_M$. The electric and magnetic polarizabilities are defined as~\cite{Bernabeu:1976jq,Llanta:1979kj,Bernard:1993bg,Bernard:1993ry}:
\begin{equation}\label{eq:def_alpha_beta}
\alpha_E+\beta_M  =-\left.\frac{1}{8 \pi} \frac{\partial^2}{\partial \omega^2} U(\omega)\right|_{\omega=0},\qquad
\beta_M  =-\frac{1}{4 \pi} V(\omega=0).
\end{equation}
Our subsequent task is to calculate all contributions to $U(\omega)$ and $V(\omega)$ up to $\mathcal{O}(p^3)$ within HB$\chi$PT.

\subsection{Effective Lagrangians}\label{subsec:effective_Lagrangian}
We choose the nonlinear realization of the chiral symmetry,
\begin{equation}\label{eq:nonlinear_realization}
	U=u^2=\exp \left(i \frac{\phi}{F_0}\right),
\end{equation}
where $\phi$ is the matrix for octet Goldstones,
\begin{equation}\label{eq:octet_meson}
\phi=\sum_{a=1}^{8}\lambda_a\phi_a=\left(\begin{array}{ccc}
\pi^0+\frac{1}{\sqrt{3}} \eta & \sqrt{2} \pi^{+} & \sqrt{2} K^{+} \\
\sqrt{2} \pi^{-} & -\pi^0+\frac{1}{\sqrt{3}} \eta & \sqrt{2} K^0 \\
\sqrt{2} K^{-} & \sqrt{2} \bar{K}^0 & -\frac{2}{\sqrt{3}} \eta
\end{array}\right),
\end{equation}
and $F_0$ is the decay constant of the pseudoscalar meson in chiral limit. We adopt $F_\pi=92.4~ \mathrm{MeV}, F_K=113~\mathrm{MeV}$ and $F_\eta=116~\mathrm{MeV}$ in this work. Under the $\mathrm{SU}(3)_L \times \mathrm{SU}(3)_R$ chiral transformation, the $U,~u$ in Eq.~\eqref{eq:nonlinear_realization} and $\mathcal{B}_n$ in Eq.~\eqref{eq:heavy_baryon_field} are transformed as follows:
\begin{equation}
\begin{aligned}
U & \rightarrow R U L^{\dagger}, \\
u & \rightarrow R u K^{\dagger}=K u L^{\dagger},\\
\mathcal{B}_n & \rightarrow K \mathcal{B}_n K^T,
\end{aligned}
\end{equation}
where $R$ and $L$ are $\mathrm{SU}(3)_R$ and $\mathrm{SU}(3)_L$ transformation matrices, respectively. $K=K(R, L, \phi)$ is a unitary transformation.

The electromagnetic fields are introduced as the left-handed and the right-handed external fields:
\begin{equation}
	r_\mu=l_\mu=-e Q_{m(\mathcal{B})} A_\mu,
\end{equation}
where $A_\mu$ is the electromagnetic field and $Q_{m(\mathcal{B})}$ represents the charge matrix. $Q_m = \operatorname{diag}\left(\frac{2}{3}, -\frac{1}{3}, -\frac{1}{3}\right)$ and $Q_{\mathcal{B}} = \operatorname{diag}(1, 0, 0)$ represent the charge matrices for light Goldstone mesons and singly charmed baryons, respectively.

We can define some ``building blocks'' of the Lagrangian in advance. The spin matrix of the heavy-baryon is defined as
\begin{equation}
	S^\mu=\frac{i}{2} \gamma_5 \sigma^{\mu \nu} v_\nu=-\frac{1}{2} \gamma_5\left(\gamma^\mu \slashed v-v^\mu\right), \quad S^{\mu \dagger}=\gamma_0 S^\mu \gamma_0.
\end{equation}
The chiral connection and vielbein are defined as \cite{Bernard:1995dp,Scherer:2002tk}
\begin{equation}
	\Gamma_\mu=\frac{1}{2}\left[u^{\dagger}\left(\partial_\mu-i r_\mu\right) u+u\left(\partial_\mu-i l_\mu\right) u^{\dagger}\right],
\end{equation}
\begin{equation}
	u_\mu=\frac{i}{2}\left[u^{\dagger}\left(\partial_\mu-i r_\mu\right) u-u\left(\partial_\mu-i l_\mu\right) u^{\dagger}\right].
\end{equation}
The covariant derivatives of the Goldstone fields and baryon fields are defined as
\begin{equation}
\begin{aligned}
	\nabla_\mu U&=\partial_\mu U-i r_\mu U+i U l_\mu,\\
	D_\mu \mathcal{B}&=\partial_\mu \mathcal{B}+\Gamma_\mu \mathcal{B}+\mathcal{B} \Gamma_\mu^{\mathrm{T}}.
\end{aligned}
\end{equation}
The chiral covariant electromagnetic field strength tensors $F_{\mu \nu}^{ \pm}$are defined as
\begin{equation}
\begin{aligned}
F_{\mu \nu}^{ \pm} & =u^{\dagger} F_{\mu \nu}^R u \pm u F_{\mu \nu}^L u^{\dagger}, \\
F_{\mu \nu}^R & =\partial_\mu r_\nu-\partial_\nu r_\mu-i\left[r_\mu, r_\nu\right], \\
F_{\mu \nu}^L & =\partial_\mu l_\nu-\partial_\nu l_\mu-i\left[l_\mu, l_\nu\right].
\end{aligned}
\end{equation}
In order to introduce the chiral symmetry breaking effect, we define $\chi_{ \pm}$,
\begin{equation}
	\begin{gathered}
\chi_{ \pm}=u^{\dagger} \chi u^{\dagger} \pm u \chi^{\dagger} u, \\
\chi=2 B_0 \operatorname{diag}\left(m_u, m_d, m_s\right).
\end{gathered}
\end{equation}
where $B_0$ is a parameter related to the quark condensate and $m_{u, d, s}$ is the current quark mass.

The leading order (LO) pure-meson Lagrangian is
\begin{equation}
	\mathcal{L}_{\phi \phi}^{(2)}=\frac{F_0^2}{4}\left\langle\nabla_\mu U\left(\nabla^\mu U\right)^{\dagger}\right\rangle+\frac{F_0^2}{4}\left\langle\chi U+ U\chi^\dagger\right\rangle,
\end{equation}
where the superscript denotes the chiral order. The $\langle \cdots \rangle$ denotes the trace in the flavor space.

In the framework of HB$\chi$PT , the LO heavy baryon Lagrangian reads
\begin{equation}\label{eq:L_BPhi_1}
\begin{aligned}
\mathcal{L}_{\mathcal{B} \phi}^{(1)} = &\frac{1}{2}\left\langle\bar{\mathcal{B}}_{\bar{3}} i v \cdot D \mathcal{B}_{\bar{3}}\right\rangle+\left\langle\bar{\mathcal{B}}_6\left(i v \cdot D-\delta_2\right) \mathcal{B}_6\right\rangle-\left\langle\bar{\mathcal{B}}_6^*\left(i v \cdot D-\delta_3\right) \mathcal{B}_6^*\right\rangle \\
& +2 g_1\left\langle\bar{\mathcal{B}}_6 S \cdot u \mathcal{B}_6\right\rangle+2 g_2\left\langle\bar{\mathcal{B}}_6 S \cdot u \mathcal{B}_{\bar{3}}+\text { H.c. }\right\rangle+g_3\left\langle\bar{\mathcal{B}}_{6 \mu}^* u^\mu \mathcal{B}_6+\text { H.c. }\right\rangle \\
& +g_4\left\langle\bar{\mathcal{B}}_{6 \mu}^* u^\mu \mathcal{B}_{\bar{3}}+\text { H.c. }\right\rangle+2 g_5\left\langle\bar{\mathcal{B}}_6^* S \cdot u \mathcal{B}_6^*\right\rangle+2 g_6\left\langle\bar{\mathcal{B}}_{\bar{3}} S \cdot u \mathcal{B}_{\bar{3}}\right\rangle,
\end{aligned}
\end{equation}
where we ignore the terms suppressed by $\frac{1}{M_{\mathcal{B}}}$. $\delta_{1,2,3}$ are the mass differences between different multiplets,
\begin{equation}
\begin{aligned}
	&\delta_1=M_{\mathcal{B}_{6}^*}-M_{\mathcal{B}_{6}}= 67~\mathrm{MeV}, \\
	&\delta_2=M_{\mathcal{B}_{6}}-M_{\mathcal{B}_{\bar{3}}}= 127~\mathrm{MeV}, \\
	&\delta_3=M_{\mathcal{B}_{6}^*}-M_{\mathcal{B}_{\bar{3}}}= 194 ~\mathrm{MeV}.
\end{aligned}
\end{equation}
$M_{\mathcal{B}_{\bar{3}}}, M_{\mathcal{B}_{6}}$ and $M_{\mathcal{B}_{6}^*}$ are the average baryon masses for the antitriplet, spin-$\frac{1}{2}$ sextet and spin-$\frac{3}{2}$ sextet, respectively. In this work, we ignore the mass splitting among the particles in the same multiplet. $g_i$ is the coupling for the interaction between the pseudoscalar mesons and heavy baryons. $g_6$ is the coupling constant between pseudoscalar mesons and antitriplet heavy baryons. The light $\operatorname{spin} S_l=0$ for the antitriplets. The pseudoscalar mesons only interact with the light degree in the heavy baryon. Thus, the parity and angular momentum conservation forbid the $\phi \mathcal{B}_{\bar {3}} \mathcal{B}_{\bar{3}}$ vertex and $g_6=0$.

The next-to-leading order (NLO) heavy baryon Lagrangian reads:
\begin{equation}\label{eq:L_BPhi_2}
\begin{aligned}
\mathcal{L}_{\mathcal{B} \phi}^{(2)} = & \frac{1}{2}\left\langle \bar{\mathcal{B}}_{\bar{3}} \frac{\left(v\cdot D\right)^2-D^2}{2M_{\bar{3}}}\mathcal{B}_{\bar{3}}\right\rangle+\left\langle \bar{\mathcal{B}}_6 \frac{\left(v\cdot D\right)^2-D^2}{2M_{6}}\mathcal{B}_6\right\rangle-\left\langle \bar {\mathcal{B}}_6^{*\mu} \frac{\left(v\cdot D\right)^2-D^2}{2M_{6^*}}\mathcal{B}_{6\mu}^*\right\rangle\\
& -\frac{i d_2}{4 M_N}\left\langle\bar{\mathcal{B}}_3\left[S^\mu, S^\nu\right] \hat{F}_{\mu \nu}^{+} \mathcal{B}_3\right\rangle-\frac{i d_3}{4 M_N}\left\langle\bar{\mathcal{B}}_3\left[S^\mu, S^\nu\right] \mathcal{B}_3\right\rangle\left\langle F_{\mu \nu}^{+}\right\rangle-\frac{i d_5}{4 M_N}\left\langle\bar{\mathcal{B}}_6\left[S^\mu, S^\nu\right] \hat{F}_{\mu \nu}^{+} \mathcal{B}_6\right\rangle \\
& -\frac{i d_6}{4 M_N}\left\langle\bar{\mathcal{B}}_6\left[S^\mu, S^\nu\right] \mathcal{B}_6\right\rangle\left\langle F_{\mu \nu}^{+}\right\rangle-\frac{i f_2}{4 M_N}\left\langle\bar{\mathcal{B}}_3\left[S^\mu, S^\nu\right] \hat{F}_{\mu \nu}^{+} \mathcal{B}_6\right\rangle+\text { H.c. }+\frac{i f_4}{4 M_N}\left\langle\bar{\mathcal{B}}_3 \hat{F}_{\mu \nu}^{+} S^\nu \mathcal{B}_6^{* \mu}\right\rangle+\text { H.c. } \\
& +\frac{i f_6}{4 M_N}\left\langle\bar{\mathcal{B}}_6 \hat{F}_{\mu \nu}^{+} S^\nu \mathcal{B}_6^{* \mu}\right\rangle+\text { H.c. }+\frac{i f_7}{4 M_N}\left\langle\bar{\mathcal{B}}_6 S^\nu \mathcal{B}_6^{* \mu}\right\rangle\left\langle F_{\mu \nu}^{+}\right\rangle+\text {H.c. }+\frac{i f_9}{4 M_N}\left\langle\bar{\mathcal{B}}_6^{* \mu} \hat{F}_{\mu \nu}^{+} \mathcal{B}_6^{* \nu}\right\rangle \\
& +\frac{i f_{10}}{4 M_N}\left\langle\bar{\mathcal{B}}_6^{* \mu} \mathcal{B}_6^{* \nu}\right\rangle\left\langle F_{\mu \nu}^{+}\right\rangle,
\end{aligned}
\end{equation}
where $d_i$ and $f_i$ are the coupling constants. $\hat{f}_{\mu \nu}^{+}=f_{\mu \nu}^{+}-\frac{1}{3} \left\langle f_{\mu \nu}^{+}\right\rangle$ is related to the traceless charge matrix of the light quarks $Q_l=\operatorname{diag}\left(\frac{2}{3},-\frac{1}{3},-\frac{1}{3}\right)$, and $\left\langle f_{\mu \nu}^{+}\right\rangle$ is related to the charge matrix of the charm quark $Q_c=\operatorname{diag}\left(\frac{1}{3}, \frac{1}{3}, \frac{1}{3}\right)$. We use the nucleon mass to render the LECs dimensionless, even though it is an irrelevant scale for the current calculation. This choice simplifies the numerical computations, particularly since many inputs, such as magnetic moments, are expressed in units of the nuclear magneton. In the following Feynman diagrams, we use solid dots (``{\large $\bullet$}'') to represent the vertices from $\mathcal{L}_{\mathcal{B} \phi}^{(2)}$.

The $\mathcal{L}_{\phi \mathcal{B}}^{(3)}$ leads to a Compton scattering amplitude that is odd in the photon momentum. However, the forward Compton scattering is even in the photon momentum due to the crossing symmetry. Thus $\mathcal{L}_{\phi \mathcal{B}}^{(3)}$ does not contribute to the polarizabilities and does not need to be considered explicitly~\cite{Bernard:1991rq,Bernard:1991ru,Bernard:1992qa,Bernard:1993bg,Bernard:1993ry}. \clabel[WZWterm]{Also, the anomalous $\pi^0 \rightarrow \gamma \gamma$ vertex leads to an amplitude which is proportional to $\epsilon^{\mu\nu\rho\sigma}k_\mu\epsilon_\nu k_\rho^\prime \epsilon^\prime_\sigma$ and does not contribute in the forward direction.}

According to the standard power counting~\cite{Bernard:1995dp,Scherer:2002tk}, the chiral order $D_\chi$ of a Feynman diagram is
\begin{equation}
	D_{\chi}=2L+1+\sum_d (d-2)N_d^{\phi}+\sum_d(d-1)N_d^{\phi B},
\end{equation}
where $L$, $N_d^{\phi}$ and $N_d^{\phi B}$ are the numbers of loops,
pure meson vertices and meson-baryon vertices, respectively. $d$ is
the chiral dimension. 

\begin{figure}[htbp]
\centering
\includegraphics[width=1.0\linewidth]{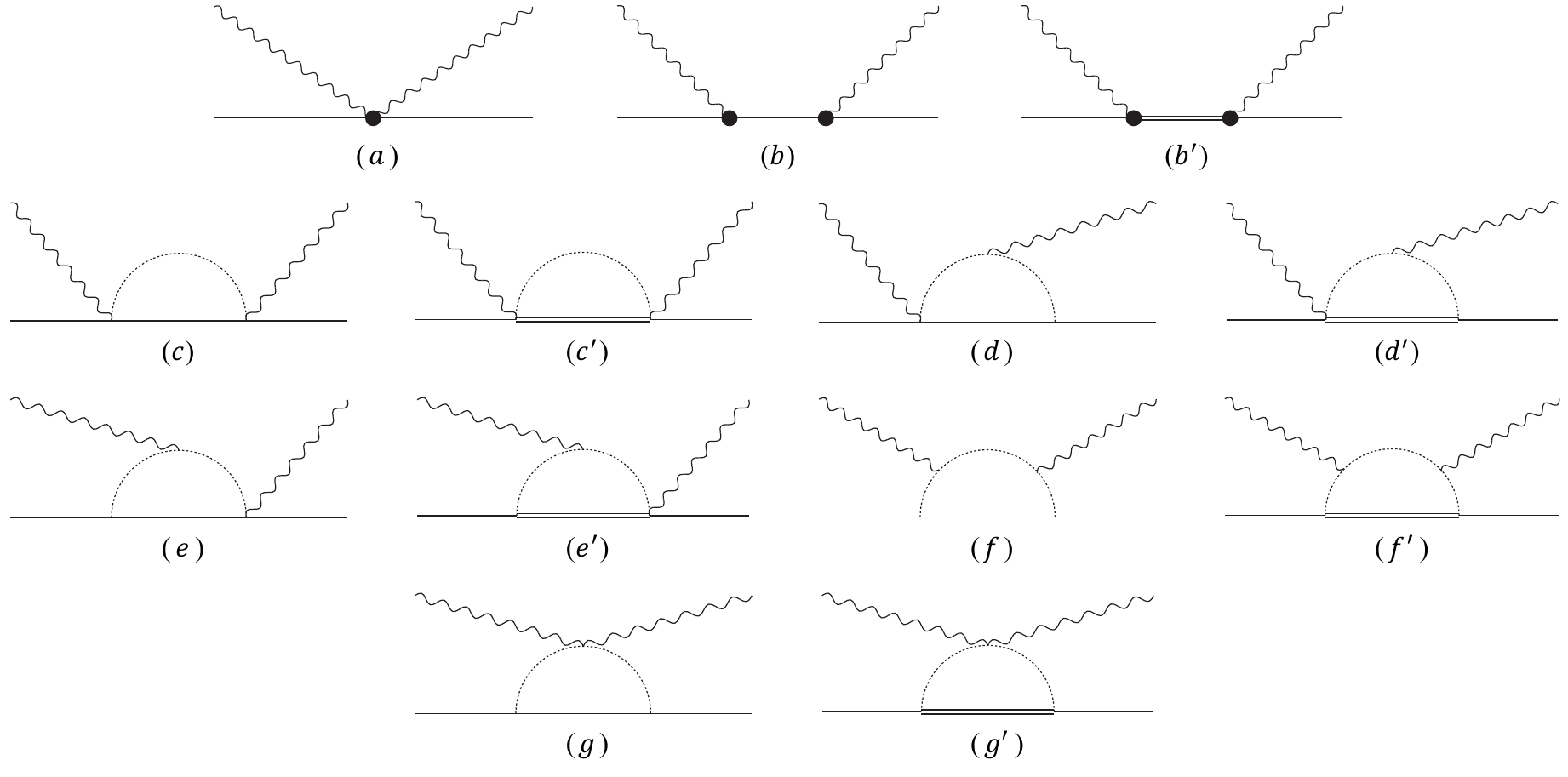}
\caption{The tree and loop diagrams contributing to the electromagnetic polarizabilities up to $\mathcal{O}\left(p^3\right)$. The solid dots (“{\large $\bullet$}”) denote the $\mathcal{L}_{\phi \mathcal{B}}^{(2)}$ vertices. The single and double lines represent the spin-$\frac{1}{2}$ and spin-$\frac{3}{2}$heavy baryons, respectively. Crossed diagrams are not shown.}
\label{fig:fmdiagrams_full}
\end{figure}

In the ``Coulomb gauge," the \(\gamma \mathcal{B}\mathcal{B}\) vertex derived from \(\mathcal{L}_{\phi \mathcal{B}}^{(1)}\) is proportional to \(\epsilon \cdot v = 0\), significantly reducing the number of diagrams that need to be calculated. The tree and loop Feynman diagrams contributing to the electromagnetic polarizabilities up to \(\mathcal{O}(p^3)\) are shown in Fig.~\ref{fig:fmdiagrams_full}.


\section{ANALYTICAL EXPRESSIONS}\label{sec:ANALYTICAL_EXPRESSIONS}

In Refs.~\cite{Meng:2018gan,Wang:2018gpl,Wang:2018cre}, the authors calculate the magnetic moments of singly heavy baryons using HB$\chi$PT. These calculations show that, due to the large mass splitting $\delta_{2,3}$, decoupling $\mathcal{B}_{\bar{3}}$ from $\mathcal{B}_{6}^{(*)}$ improves chiral convergence. In contrast, \(\mathcal{B}_{6}\) and \(\mathcal{B}_6^*\) are nearly degenerate in the heavy quark limit, and the smaller mass splitting \(\delta_1\) does not significantly affect chiral convergence. Furthermore, when accounting for chiral fluctuations, it is crucial to include the coupling between \(\mathcal{B}_{6}\) and \(\mathcal{B}_6^*\), as required by heavy quark spin symmetry. In the following discussion, we adopt this approach.

\subsection{Electromagnetic polarizabilities of $\mathcal{B}_{\bar{3}}$}
Since $\mathcal{B}_{\bar{3}}$ and $\mathcal{B}_6^{(*)}$ are decoupled, and the $\phi \mathcal{B}_3 \mathcal{B}_3$ vertex is forbidden, the spin-averaged forward Compton amplitudes of $\mathcal{B}_{\bar{3}}$ only include contributions from tree diagrams in Figs.~\ref{fig:fmdiagrams_full}($a$) and ($b$), which gives
\begin{equation}\label{eq:UV_B3}
	 U_{\xi}(\omega)=\frac{Q_\xi^2e^2}{M_{\bar{3}}},\qquad V_{\xi}(\omega)=0,
\end{equation}
where $\xi$ denotes a specific baryon in the antitriplet of Eq.~\eqref{eq:baryon_multi}, and $Q_\xi$ represents its electric charge. Equation~\eqref{eq:UV_B3} is consistent with the well-known Thomson amplitude:
	\begin{equation}
	\mathcal{M}_{\text{Thomson}}=-\frac{Q^2}{M} \bm{\epsilon} \cdot \bm{\epsilon}^{\prime}.
\end{equation}
The electromagnetic polarizabilities of $\mathcal{B}_{\bar{3}}$ are
\begin{equation}
\begin{aligned}
	&\alpha_E(\Lambda_c^+)=\alpha_E(\Xi_c^+)=\alpha_E(\Xi_c^0)=0,\\
	&\beta_M(\Lambda_c^+)=\beta_M(\Xi_c^+)=\beta_M(\Xi_c^0)=0.
\end{aligned}
\end{equation}
The vanishing electromagnetic polarizabilities of $\mathcal{B}_{\bar{3}}$ at this order is actually quite reasonable. The parity and angular momentum conservation forbid the chiral interactions for $\mathcal{B}_{\bar{3}}$. Therefore, $\mathcal{B}_{\bar{3}}$ behaves like a charged point particle without any electromagnetic polarizability up to $\mathcal{O}(p^3)$.

\subsection{Electromagnetic polarizabilities of $\mathcal{B}_{6}$}
The spin-averaged forward Compton amplitudes of $\mathcal{B}_{6}$ include contributions from both tree diagrams and $\phi \mathcal{B}$-loop diagrams. Additionally, the coupling between $\mathcal{B}_6$ and $\mathcal{B}_6^*$ allows the spin-$\frac{3}{2}$ baryons to appear as intermediate states. Here we use $\xi$ to denote a specific baryon in the spin-$\frac{1}{2}$ sextet of Eq.~\eqref{eq:baryon_multi}.
\subsubsection{Tree diagrams}
The tree diagrams in Figs.~\ref{fig:fmdiagrams_full}($a$) and ($b$) yield the Thomson amplitude
\begin{equation}\label{eq:UV_B6_ab}
	U_{\xi}^{(a+b)}(\omega)=\frac{Q_\xi^2 e^2}{M_{\mathcal{B}_6}}, \quad V_{\xi}^{(a+b)}(\omega)=0,
\end{equation}
which does not contribute to the electromagnetic polarizabilities:
\begin{equation}
	\alpha_E^{(a+b)}\left(\xi\right)=\beta_M^{(a+b)}\left(\xi\right)=0.
\end{equation}

The tree diagram with $\mathcal{B}_6^*$ as the intermediate state, shown in Fig.~\ref{fig:fmdiagrams_full}($b^\prime$), yields:
\begin{equation}\label{eq:UV_fig_b_2}
	U_{\xi}^{(b^\prime)}(\omega)=-\frac{e^2 C_\xi^2\omega^2}{12 M_N^2}\frac{\delta_1}{\delta_1^2-\omega^2},\quad
	V_{\xi}^{(b^\prime)}(\omega)=-\frac{e^2 C_\xi^2}{12 M_N^2}\frac{\delta_1}{\delta_1^2-\omega^2},
\end{equation}
where $C_\xi$ represents the coefficients for different baryons, as listed in Table~\ref{tab:C_xi}. The electromagnetic polarizabilities from Eq.~\eqref{eq:UV_fig_b_2} are:
\begin{equation}\label{eq:ab_b6_b2}
	\alpha^{(b^\prime)}_{E}(\xi)=0,\quad \beta^{(b^\prime)}_{M}(\xi)=\frac{\alpha_{\mathrm{em}} C_\xi^2}{12 M_N^2} \frac{1}{\delta_1}.
\end{equation}
We observe that Fig.~\ref{fig:fmdiagrams_full}($b^\prime$) does not contribute to the electric polarizability but introduces a magnetic polarizability with a $\delta_1$ pole term. A similar effect occurs when considering the contribution of $\Delta$-baryons to nucleon polarizabilities, as discussed in Refs.~\cite{Schumacher:2005an,Holstein:2013kia,Hemmert:1996rw,Pascalutsa:2002pi}. The only difference is that the mass splitting $\delta_1$ is even smaller than $M_\Delta - M_N$ and approaches zero in the heavy quark limit. As a result, Eq.~\eqref{eq:ab_b6_b2} provides a sizable contribution, which diverges in the heavy quark limit due to the degenerate $\mathcal{B}_6$ and $\mathcal{B}_6^*$. At the same time, Fig.~\ref{fig:fmdiagrams_full}($b^\prime$) purely represents two consecutive magnetic dipole transitions, so it does not contribute to the electric polarizability. 

We summarize the (transition) magnetic dipole moments obtained from the quark model~\cite{Wang:2018gpl}, leading order HB$\chi$PT calculations~\cite{Meng:2018gan}, and lattice QCD simulations~\cite{Can:2013tna,Bahtiyar:2016dom,Can:2015exa} in Tables~\ref{tab:magnetic_moments_B3} and \ref{tab:magnetic_moments}. The coefficient $C_\xi$ is proportional to the transition magnetic moment from the LO result in HB$\chi$PT:
\begin{equation}\label{eq:transition_LO}
	C_\xi=-\sqrt{6}~\mu^{\mathrm{LO}}_{\xi^*\to \xi +\gamma}.
\end{equation}

\begin{table}[htbp]
    \centering
    \caption{The coefficients $C_\xi$ of the diagram in Fig.~\ref{fig:fmdiagrams_full}($b^\prime$)}
    \label{tab:C_xi}
    \setlength{\tabcolsep}{2.5mm}
    \begin{tabular}{c|cccccc}
    \hline\hline
         & $\Sigma_c^{++}$ & $\Sigma_c^{+}$ & $\Sigma_c^{0}$ & $\Xi_{c}^{\prime+}$ & $\Xi_{c}^{\prime0}$ & $\Omega_c^{0}$  \\ \hline
        $C$ & $f_7+\frac{2}{3}f_6$ & $f_7+\frac{1}{6}f_6$ & $f_7-\frac{1}{3}f_6$ & $f_7+\frac{1}{6}f_6$ & $f_7-\frac{1}{3}f_6$ & $f_7-\frac{1}{3}f_6$  \\
        \hline\hline
    \end{tabular}
\end{table}

\begin{table}[htbp]
    \centering
    \caption{The magnetic moments $\mu_{\mathcal{B}_3}$ from the quark model (QM)~\cite{Wang:2018gpl}, leading order HB$\chi$PT calculations~\cite{Meng:2018gan}, and lattice QCD simulations~\cite{Can:2013tna,Bahtiyar:2016dom,Can:2015exa}. Magnetic moments are given in units of the nuclear magneton $\mu_N$.}
    \label{tab:magnetic_moments_B3}
    \setlength{\tabcolsep}{3mm}
    \begin{tabular}{c|c|c|c}
    \hline\hline
  $~$  & $\Lambda_c^+$ & $\Xi_c^+$ & $\Xi_c^0$  \\
\hline
QM & $\mu_c$ & $\mu_c$ & $\mu_c$ \\
HB$\chi$PT & $\frac{1}{3}d_2+2d_3$ & $\frac{1}{3}d_2+2d_3$ & $-\frac{2}{3}d_2+2d_3$\\
LQCD & ~ & $0.235(25)$ & $0.192(17)$\\
\hline\hline
    \end{tabular}
\end{table}

\begin{table}[htbp]
\centering
\caption{The (transition) magnetic moments $\mu_\xi$ and $\mu_{\xi^*\to \xi +\gamma}$ from the quark model (QM)~\cite{Wang:2018gpl}, leading order HB$\chi$PT calculations~\cite{Meng:2018gan}, and lattice QCD simulations~\cite{Can:2013tna,Bahtiyar:2016dom,Can:2015exa}. Magnetic moments are given in units of the nuclear magneton $\mu_N$.}
\label{tab:magnetic_moments}
\begin{tabular}{c|c|c|c|c|c|c}
\hline \hline
~ & $\Sigma_c^{++}$ & $\Sigma_c^{+}$ & $\Sigma_c^0$ & $\Xi_c^{\prime+}$ & $\Xi_c^{\prime 0}$ & $\Omega_c^0$ \\
\hline QM & $\frac{4}{3} \mu_u-\frac{1}{3} \mu_c$ & $\frac{2}{3} \mu_u+\frac{2}{3} \mu_d-\frac{1}{3} \mu_c$ & $\frac{4}{3} \mu_d-\frac{1}{3} \mu_c$ & $\frac{2}{3} \mu_u+\frac{2}{3} \mu_s-\frac{1}{3} \mu_c$ & $\frac{2}{3} \mu_d+\frac{2}{3} \mu_s-\frac{1}{3} \mu_c$ & $\frac{4}{3} \mu_s-\frac{1}{3} \mu_c$ \\
HB$\chi$PT & $\frac{2}{3} d_5+d_6$ & $\frac{1}{6} d_5+d_6$ & $-\frac{1}{3} d_5+d_6$ & $\frac{1}{6} d_5+d_6$ & $-\frac{1}{3} d_5+d_6$ & $-\frac{1}{3} d_5+d_6$ \\
LQCD & $1.499(202)$ & $~$ & $-0.875(103)$ & $0.315(141)$ & $-0.599(71)$ &$-0.688(31)$\\
\hline
~ & $\Sigma_c^{*++} \rightarrow \Sigma_c^{++} \gamma$ & $\Sigma_c^{*+} \rightarrow \Sigma_c^{+} \gamma$ & $\Sigma_c^{* 0} \rightarrow \Sigma_c^0 \gamma$ & $\Xi_c^{*+} \rightarrow \Xi_c^{\prime+} \gamma$ & $\Xi_c^{* 0} \rightarrow \Xi_c^{\prime 0} \gamma$ & $\Omega_c^{* 0} \rightarrow \Omega^0 \gamma$ \\
\hline QM & $\frac{2 \sqrt{2}}{3}\left(\mu_u-\mu_c\right)$ & $\frac{\sqrt{2}}{3}\left(\mu_u+\mu_d-2 \mu_c\right)$ & $\frac{2 \sqrt{2}}{3}\left(\mu_d-\mu_c\right)$ & $\frac{\sqrt{2}}{3}\left(\mu_u+\mu_s-2 \mu_c\right)$ & $\frac{\sqrt{2}}{3}\left(\mu_d+\mu_s-2 \mu_c\right)$ & $\frac{2 \sqrt{2}}{3}\left(\mu_s-\mu_c\right)$ \\
HB$\chi$PT & $-\sqrt{\frac{1}{6}}\left(\frac{2}{3} f_6+f_7\right)$ & $-\sqrt{\frac{1}{6}}\left(\frac{1}{6} f_6+f_7\right)$ & $-\sqrt{\frac{1}{6}}\left(-\frac{1}{3} f_6+f_7\right)$ & $-\sqrt{\frac{1}{6}}\left(\frac{1}{6} f_6+f_7\right)$ & $-\sqrt{\frac{1}{6}}\left(-\frac{1}{3} f_6+f_7\right)$ & $-\sqrt{\frac{1}{6}}\left(-\frac{1}{3} f_6+f_7\right)$ \\
\hline \hline
\end{tabular}
\end{table}

\subsubsection{Loop diagrams}

Using the $\mathcal{J}$-function defined in Appendix~\ref{appendix:loop_integrals}, we can obtain the form factors of the $\mathcal{B}_6\phi$-loop diagrams in Figs.~\ref{fig:fmdiagrams_full}($c$)--($g$):
\begin{align}
U^{(c)}_{\xi}(\omega)= & A\sum_{\chi}\frac{D^{(c)}_{\xi,\chi}}{F_\chi^2}\mathcal{J}_0(\omega,0,M_{\chi}^2),\label{eq:U_6_c}\\
	U^{(d+e)}_{\xi}(\omega)= & A\left[\sum_{\chi} \frac{D_{\xi,\chi}^{(d+e)}}{F_\chi^2}\int_0^1 dx~\mathcal{J}^{\prime}_2(\omega x,0,M_{\chi}^2)\right],\\
	U^{(f)}_{\xi}(\omega)= & A\left[\sum_{\chi} \frac{D_{\xi,\chi}^{(f)}}{F_\chi^2}\int_0^1dx~ (1-x)(d+1)\mathcal{J}_6^{\prime\prime}(\omega x,0,M_{\chi}^2)\right.\notag \\
	&\quad ~ -\left. \sum_{\chi} \frac{D_{\xi,\chi}^{(f)}}{F_\chi^2} \int_0^1dx~\omega^2(1-x)x^2\mathcal{J}_2^{\prime\prime}(\omega x,0,M_{\chi}^2)\right],\\
	U^{(g)}_{\xi}(\omega)= &A\sum_{\chi}\frac{D^{(g)}_{\xi,\chi}}{F_\chi^2}(d-1)\mathcal{J}_2^{\prime}(0,0,M_{\chi}^2),\\
	V^{(c)}_{\xi}(\omega)= & 0,\\
	V^{(d+e)}_{\xi}(\omega)= & A\left[\sum_{\chi} -\frac{1}{2 F_\chi^2}D_{\xi,\chi}^{(d+e)}\int_0^1 dx~x(1-2x)\mathcal{J}^{\prime}_0(\omega x,0,M_{\chi}^2)\right],\\
	V^{(f)}_{\xi}(\omega)= & A\left[\sum_{\chi} \frac{1}{4 F_\chi^2}D^{(f)}_{\xi,\chi}\int_0^1dx~(1-x)\left[8x(2x-1)+(2x-1)^2(d-1)\right]\mathcal{J}_2^{\prime\prime}(\omega x,0,M_{\chi}^2)\right.\notag \\
	 &\quad ~ -\left. \sum_{\chi} \frac{1}{4 F_\chi^2}D^{(f)}_{\xi,\chi} \int_0^1 dx~ \omega^2(1-x)x^2(2x-1)^2 \mathcal{J}_0^{\prime\prime}(\omega x, 0,M_{\chi}^2)\right],\\
	 V^{(g)}_{\xi}(\omega)= & 0.\label{eq:V_6_g}
\end{align}
In Eqs.~\eqref{eq:U_6_c}--\eqref{eq:V_6_g}, we have considered the contributions from the crossed diagrams. $\chi$ denotes a specific meson in Eq.~\eqref{eq:octet_meson}, and $M_\chi$ represents its mass. $D_{\xi,\chi}$ are the coefficients of the loops, which are given in Table~\ref{tab:loop_cofficients}. $A$ is a common factor
\begin{equation}
	A=i\frac{e^2g_1^2}{2}\,.
\end{equation}
Expanding Eqs.~\eqref{eq:U_6_c}--\eqref{eq:V_6_g} into a power series in $\omega$ and combining with Eq.~\eqref{eq:def_alpha_beta}, we can obtain the electromagnetic polarizabilities from the $\mathcal{B}_6\phi$-loop diagrams in Figs.~\ref{fig:fmdiagrams_full}($c$)--($g$):
\begin{align}
	\alpha_E^{(c-g)}(\Sigma_c^{++})=& \frac{5 \alpha_{\mathrm{em}}g_1^2}{192 \pi M_\pi F_\pi^2}+ \frac{5 \alpha_{\mathrm{em}}g_1^2}{192 \pi M_K F_K^2},\\
	\alpha_E^{(c-g)}(\Sigma_c^+)=& \frac{10 \alpha_{\mathrm{em}}g_1^2}{192 \pi M_\pi F_\pi^2}+ \frac{5 \alpha_{\mathrm{em}}g_1^2}{384 \pi M_K F_K^2},\\
	\alpha_E^{(c-g)}(\Sigma_c^0)=& \frac{5 \alpha_{\mathrm{em}}g_1^2}{192 \pi M_\pi F_\pi^2},\\
	\alpha_E^{(c-g)}(\Xi_c^{\prime +})=& \frac{5 \alpha_{\mathrm{em}}g_1^2}{384 \pi M_\pi F_\pi^2}+ \frac{10 \alpha_{\mathrm{em}}g_1^2}{192 \pi M_K F_K^2},\\
	\alpha_E^{(c-g)}(\Xi_c^{\prime 0})=& \frac{5 \alpha_{\mathrm{em}}g_1^2}{384 \pi M_\pi F_\pi^2}+ \frac{5 \alpha_{\mathrm{em}}g_1^2}{384 \pi M_K F_K^2},\\
	\alpha_E^{(c-g)}(\Omega_c^0)=& \frac{5 \alpha_{\mathrm{em}}g_1^2}{192 \pi M_K F_K^2},\\
	\beta_M^{(c-g)}(\xi) =& \frac{1}{10} \alpha_E^{(c-g)}(\xi).
\end{align}
The results diverge as $1/m_\chi$ in the chiral limit, similar to the behavior of nucleon polarizabilities discussed in Refs.~\cite{Bernard:1991rq,Bernard:1991ru,Bernard:1993bg,Bernard:1993ry}.  These results indicate that the photons interact intensively with the meson cloud at low energies and that the long-range chiral corrections contribute significantly to the electromagnetic polarizabilities.

\begin{table}[htbp]
    \centering
    \caption{The coefficients of the loop diagrams in Fig.~\ref{fig:fmdiagrams_full}.}
    \label{tab:loop_cofficients}
    \setlength{\tabcolsep}{3mm}
    \begin{tabular}{c|cccccc}
    \hline\hline
         & $\Sigma_c^{++}$ & $\Sigma_c^{+}$ & $\Sigma_c^{0}$ & $\Xi_{c}^{\prime+}$ & $\Xi_{c}^{\prime0}$ & $\Omega_c^{0}$  \\ \hline
        $D_{\pi}^{(c)}$ & $-\frac{1}{2}$ & $-1$ & $-\frac{1}{2}$ & $-\frac{1}{4}$ & $-\frac{1}{4}$ & $0$  \\ 
        $D_{K}^{(c)}$ & $-\frac{1}{2}$ & $-\frac{1}{4}$ & $0$ & $-1$ & $-\frac{1}{4}$ & $-\frac{1}{2}$  \\
        $D_{\pi}^{(d+e)}$ & $2$ & $4$ & $2$ & $1$ & $1$ & $0$  \\ 
        $D_{K}^{(d+e)}$ & $2$ & $1$ & $0$ & $4$ & $1$ & $2$  \\ 
        $D_{\pi}^{(f)}$ & $-2$ & $-4$ & $-2$ & $-1$ & $-1$ & $0$  \\ 
        $D_{K}^{(f)}$ & $-2$ & $-1$ & $0$ & $-4$ & $-1$ & $-2$  \\ 
        $D_{\pi}^{(g)}$ & $\frac{1}{2}$ & $1$ & $\frac{1}{2}$ & $\frac{1}{4}$ & $\frac{1}{4}$ & $0$  \\ 
        $D_{K}^{(g)}$ & $\frac{1}{2}$ & $\frac{1}{4}$ & $0$ & $1$ & $\frac{1}{4}$ & $\frac{1}{2}$ \\ 
        \hline\hline
    \end{tabular}
\end{table}

Similarly, the form factors of the $\mathcal{B}_6^*\phi$-loop diagrams in Figs.~\ref{fig:fmdiagrams_full}($c^\prime$)--($g^\prime$) are:
\begin{align}
	U^{(c^\prime)}_{\xi}(\omega)= & B\sum_{\chi}\frac{D^{(c)}_{\xi,\chi}}{F_\chi^2}\mathcal{J}_0(\omega,\delta_1,M_{\chi}^2)\label{eq:U_6s_c},\\
	U^{(d^\prime+e^\prime)}_{\xi}(\omega)= & B\left[\sum_{\chi} \frac{D_{\xi,\chi}^{(d+e)}}{F_\chi^2}\int_0^1 dx~\mathcal{J}^{\prime}_2(\omega x,\delta_1,M_{\chi}^2)\right],\\
	U^{(f^\prime)}_{\xi}(\omega)= & B\left[\sum_{\chi} \frac{D_{\xi,\chi}^{(f)}}{F_\chi^2}\int_0^1dx~ (1-x)(d+1)\mathcal{J}_6^{\prime\prime}(\omega x,\delta_1,M_{\chi}^2)\right.\notag \\
	&\quad ~ -\left. \sum_{\chi} \frac{D_{\xi,\chi}^{(f)}}{F_\chi^2} \int_0^1dx~\omega^2(1-x)x^2\mathcal{J}_2^{\prime\prime}(\omega x,\delta_1,M_{\chi}^2)\right],\\
	U^{(g^\prime)}_{\xi}(\omega)= &B\sum_{\chi}\frac{D^{(g)}_{\xi,\chi}}{F_\chi^2}(d-1)\mathcal{J}_2^{\prime}(0,\delta_1,M_{\chi}^2),\\
	V^{(c^\prime)}_{\xi}(\omega)= & 0,\\
	V^{(d^\prime+e^\prime)}_{\xi}(\omega)= & B\left[\sum_{\chi} -\frac{1}{2 F_\chi^2}D_{\xi,\chi}^{(d+e)}\int_0^1 dx~x(1-2x)\mathcal{J}^{\prime}_0(\omega x,\delta_1,M_{\chi}^2)\right],\\
	V^{(f^\prime)}_{\xi}(\omega)= & B\left[\sum_{\chi} \frac{1}{4 F_\chi^2}D^{(f)}_{\xi,\chi}\int_0^1dx~(1-x)\left[8x(2x-1)+(2x-1)^2(d-1)\right]\mathcal{J}_2^{\prime\prime}(\omega x,\delta_1,M_{\chi}^2)\right.\notag \\
	 &\quad ~ -\left. \sum_{\chi} \frac{1}{4 F_\chi^2}D^{(f)}_{\xi,\chi} \int_0^1 dx~ \omega^2(1-x)x^2(2x-1)^2 \mathcal{J}_0^{\prime\prime}(\omega x,\delta_1,M_{\chi}^2)\right],\\
	 V^{(g^\prime)}_{\xi}(\omega)= & 0,\label{eq:V_6s_g}
\end{align}
where $B$ is a common factor
\begin{equation}
	B=i\frac{e^2g_3^2}{2}\frac{d-2}{d-1}.
\end{equation}
Expanding Eq.~\eqref{eq:U_6s_c}--\eqref{eq:V_6s_g} into a power series in $\omega$ and combining with Eq.~\eqref{eq:def_alpha_beta}, we can obtain the electromagnetic polarizabilities from the $\mathcal{B}_6^*\phi$-loop diagrams in Figs.~\ref{fig:fmdiagrams_full}($c^\prime$)--($g^\prime$):
\begin{align}
	\alpha_E^{(c^\prime-g^\prime)}\left(\Sigma_c^{++}\right)=&\frac{\alpha_{\mathrm{em}} g_3^2 S_\pi}{288\pi^2 F_\pi^2 \left(M_\pi^2-\delta_1^2\right)^2}+\frac{\alpha_{\mathrm{em}} g_3^2 S_K}{288\pi^2 F_K^2 \left(M_K^2-\delta_1^2\right)^2},\\
	\alpha_E^{(c^\prime-g^\prime)}\left(\Sigma_c^{+}\right)=&\frac{\alpha_{\mathrm{em}} g_3^2 S_\pi}{144\pi^2 F_\pi^2 \left(M_\pi^2-\delta_1^2\right)^2}+\frac{\alpha_{\mathrm{em}} g_3^2 S_K}{576\pi^2 F_K^2 \left(M_K^2-\delta_1^2\right)^2},\\
	\alpha_E^{(c^\prime-g^\prime)}\left(\Sigma_c^{0}\right)=&\frac{\alpha_{\mathrm{em}} g_3^2 S_\pi}{288\pi^2 F_\pi^2 \left(M_\pi^2-\delta_1^2\right)^2},\\
	\alpha_E^{(c^\prime-g^\prime)}\left(\Xi_c^{\prime +}\right)=&\frac{\alpha_{\mathrm{em}} g_3^2 S_\pi}{576\pi^2 F_\pi^2 \left(M_\pi^2-\delta_1^2\right)^2}+\frac{\alpha_{\mathrm{em}} g_3^2 S_K}{144\pi^2 F_K^2 \left(M_K^2-\delta_1^2\right)^2},\\
	\alpha_E^{(c^\prime-g^\prime)}\left(\Xi_c^{\prime 0}\right)=&\frac{\alpha_{\mathrm{em}} g_3^2 S_\pi}{576\pi^2 F_\pi^2 \left(M_\pi^2-\delta_1^2\right)^2}+\frac{\alpha_{\mathrm{em}} g_3^2 S_K}{576\pi^2 F_K^2 \left(M_K^2-\delta_1^2\right)^2},\\
	\alpha_E^{(c^\prime-g^\prime)}\left(\Omega_c^{0}\right)=&\frac{\alpha_{\mathrm{em}} g_3^2 S_K}{288\pi^2 F_K^2 \left(M_K^2-\delta_1^2\right)^2}.\\
	\beta_M^{(c^\prime-g^\prime)}\left(\Sigma_c^{++}\right)=&\frac{\alpha_{\mathrm{em}} g_3^2 R_\pi}{288 \pi^2 F_\pi^2 (M_\pi^2-\delta_1^2)}+\frac{\alpha_{\mathrm{em}} g_3^2 R_K}{288 \pi^2 F_K^2 (M_K^2-\delta_1^2)},\\
	\beta_M^{(c^\prime-g^\prime)}\left(\Sigma_c^{+}\right)=&\frac{\alpha_{\mathrm{em}} g_3^2 R_\pi}{144 \pi^2 F_\pi^2 (M_\pi^2-\delta_1^2)}+\frac{\alpha_{\mathrm{em}} g_3^2 R_K}{576 \pi^2 F_K^2 (M_K^2-\delta_1^2)},\\
	\beta_M^{(c^\prime-g^\prime)}\left(\Sigma_c^{0}\right)=&\frac{\alpha_{\mathrm{em}} g_3^2 R_\pi}{288 \pi^2 F_\pi^2 (M_\pi^2-\delta_1^2)},\\
	\beta_M^{(c^\prime-g^\prime)}\left(\Xi_c^{\prime+}\right)=&\frac{\alpha_{\mathrm{em}} g_3^2 R_\pi}{576 \pi^2 F_\pi^2 (M_\pi^2-\delta_1^2)}+\frac{\alpha_{\mathrm{em}} g_3^2 R_K}{144 \pi^2 F_K^2 (M_K^2-\delta_1^2)},\\
	\beta_M^{(c^\prime-g^\prime)}\left(\Xi_c^{\prime 0}\right)=&\frac{\alpha_{\mathrm{em}} g_3^2 R_\pi}{576 \pi^2 F_\pi^2 (M_\pi^2-\delta_1^2)}+\frac{\alpha_{\mathrm{em}} g_3^2 R_K}{576 \pi^2 F_K^2 (M_K^2-\delta_1^2)},\\
	\beta_M^{(c^\prime-g^\prime)}\left(\Omega_c^{0}\right)=&\frac{\alpha_{\mathrm{em}} g_3^2 R_K}{288 \pi^2 F_K^2 (M_K^2-\delta_1^2)},
\end{align}
where we have defined
\begin{equation}
\begin{aligned}
	R_\chi&=\sqrt{M_\chi^2-\delta_1^2}\arccos\left[\frac{\delta_1}{M_\chi}\right],\\
	S_\chi&=M_\pi^2\left(10R_\pi-9\delta_1\right)+\delta_1^2\left(9\delta_1-R_\pi\right).
	\end{aligned}
\end{equation}

In the heavy quark limit with $g_1^2 = \frac{4}{3} g_3^2$~\cite{Yan:1992gz,Cheng:1993kp,Cho:1992nt,Jiang:2015xqa} and  $\delta_1 = 0$ , we have
\begin{equation}\label{eq:RS_HQSS}
	R_\chi = \frac{\pi M_\chi}{2}, \quad S_\chi = 5 \pi M_\chi^3.
\end{equation}
It is easy to verify that in the heavy quark limit:
\begin{equation}
	\alpha_E^{(c-g)}(\xi) = 2 \alpha_E^{(c'-g')}(\xi),\quad \beta_M^{(c-g)}(\xi) = 2 \beta_M^{(c'-g')}(\xi).
\end{equation}
In real physics, $\delta_1$ is a small nonzero mass splitting, which slightly suppresses the contribution from the $\mathcal{B}_6^*\phi$-loop diagrams. Therefore, we expect the electromagnetic polarizabilities from the $\mathcal{B}_6^*\phi$-loop diagrams should be slightly smaller than half of those from the $\mathcal{B}_6\phi$-loop diagrams.

Finally, by summing all the contributions calculated above, we can obtain the total electromagnetic polarizabilities:
\begin{equation}
	\begin{aligned}
	\alpha_E^{\mathrm{Tot.}}(\xi) &=\alpha_E^{(c-g)}(\xi)+\alpha_E^{(c^\prime-g^\prime)}(\xi),\\
	\beta_M^{\mathrm{Tot.}}(\xi) &=\beta_M^{(b^\prime)}(\xi)+\beta_M^{(c-g)}(\xi)+\beta_M^{(c^\prime-g^\prime)}(\xi).
	\end{aligned}
\end{equation}

\section{NUMERICAL RESULTS}\label{sec:NUMERICAL_RESULTS}

In the analytical expression for the polarizabilities, there are four low-energy constants (LECs) that need to be determined. These include the axial coupling constants $g_{1,3}$ from $\mathcal{L}^{(1)}_{\mathcal{B}\phi}$, and $f_{6,7}$ from $\mathcal{L}^{(2)}_{\mathcal{B}\phi}$. \clabel[g1andg2]{The coupling constants $g_2$ may be derived from the decay widths:
\begin{equation}
    \Gamma\left(\Sigma_c \rightarrow \Lambda \pi\right)=\frac{g_2^2}{4 \pi f_\pi^2} \frac{M_{\Lambda_c}}{M_{\Sigma_c}}\left|\vec{p}_\pi\right|^3
\end{equation}
where $\left|\vec{p}_\pi\right|$ is the c.m. momentum of the pion. From the decay widths of $\Sigma_c$, we obtain $\left|g_2\right|=0.568\pm0.023$. In the heavy quark limit, the heavy quark spin symmetry requires $g_1^2 = \frac{4}{3} g_3^2$~\cite{Yan:1992gz,Cheng:1993kp,Cho:1992nt,Jiang:2015xqa}. Furthermore, the quark model symmetry leads to $g_1^2 = \frac{8}{3} g_2^2$~\cite{Yan:1992gz,Liu:2011xc}. In the numerical evaluation, we use the following values:
\begin{equation}\label{eq:charm_coupling}
	g_1 = 0.928\pm0.093\pm0.037,\quad g_3 = 0.804\pm0.081\pm0.032.
\end{equation}
The first uncertainty arises from the assumed 10\% error introduced by using the quark model symmetry and the heavy quark spin symmetry, while the second one is due to the experimental uncertainty in the $\Sigma_c$ widths.} The remaining two LECs, $f_6$ and $f_7$, appear in the coefficients $C_\xi$ associated with the magnetic dipole transitions, as shown in Tables.~\ref{tab:C_xi} and~\ref{tab:magnetic_moments}. 
It is worth noting that in the quark model, there are four distinct quark magnetic moments, whereas the HB$\chi$PT results at this order include only two LECs, $f_6$ and $f_7$, without considering SU(3) flavor-breaking effects. To achieve more accurate results, we treat $\mu_u$, $\mu_d=-\frac{1}{2}\mu_u$, $\mu_s$, and $\mu_c$ as three independent parameters to account for the contributions of $f_6$ and $f_7$, additionaly with SU(3) flavor-breaking effects. We can determine the coefficient $C_\xi$ as:
\begin{equation}
	C_\xi=-\sqrt{6}~\mu_{\xi^*\to \xi +\gamma}.
\end{equation}
We fit the quark magnetic moments by minimizing
\begin{equation}
	\chi^2=\frac{1}{\mathrm{d.o.f.}}\sum_{i=1}^7\left(\frac{\mu_i^{\mathrm{QM}}-\mu_i^{\mathrm{LQCD}}}{\sigma_i^{\mathrm{LQCD}}}\right),
\end{equation}
where $\mu_i^{\mathrm{QM(LQCD)}}$ represent the baryon magnetic moments from the quark model (lattice QCD simulations) as listed in Tables~\ref{tab:magnetic_moments_B3} and~\ref{tab:magnetic_moments}, $\sigma_i^{\mathrm{LQCD}}$ is the uncertainties of $\mu_i^{\mathrm{LQCD}}$, and d.o.f refers to the degrees of freedom, given by d.o.f = 4, since the fit includes $7$ data points and $3$ free parameters. This fit yields the following results (in units of $\mu_N$):
\begin{equation}
	\mu_u=-2\mu_d=1.078(88),\quad \mu_s=-0.456(23),\quad \mu_c=0.205(15),
\end{equation}
with $\chi^2=1.71$. The values in parentheses represent the uncertainties originating from the lattice QCD results. It can be seen that the magnetic moment of the charm quark is suppressed by $1/m_c$. Then we calculate the transition magnetic moments for singly charmed baryons in quark model~\cite{Wang:2018gpl}, which are (in units of $\mu_N$):
\begin{align}
	&\mu_{\Sigma_c^{*++} \rightarrow \Sigma_c^{++} \gamma}  =\frac{2 \sqrt{2}}{3}\left(\mu_u-\mu_c\right)=0.82(10),\\
	&\mu_{\Sigma_c^{*+} \rightarrow \Sigma_c^{+} \gamma}  =\frac{ \sqrt{2}}{3}\left(\mu_u+\mu_d-2\mu_c\right)=0.06(3),\\
	&\mu_{\Sigma_c^{*0} \rightarrow \Sigma_c^{0} \gamma}  =\frac{ 2 \sqrt{2}}{3}\left(\mu_d-\mu_c\right)=-0.70(5),\\
	&\mu_{\Xi_c^{*+} \rightarrow \Xi_c^{\prime +} \gamma}  =\frac{ \sqrt{2}}{3}\left(\mu_u+\mu_s-2\mu_c\right)=0.10(6),\\
	&\mu_{\Xi_c^{*0} \rightarrow \Xi_c^{\prime 0} \gamma}  =\frac{ \sqrt{2}}{3}\left(\mu_d+\mu_s-2\mu_c\right)=-0.66(3),\\
	&\mu_{\Omega_c^{*0} \rightarrow \Omega_c^{0} \gamma}  =\frac{2 \sqrt{2}}{3}\left(\mu_s-\mu_c\right)=-0.62(3).
\end{align}

The numerical results for the electromagnetic polarizabilities of singly charmed baryons are listed in Table.~\ref{tab:polarizabilities_numerical_results}. We observe that
\begin{equation}
	\alpha_E^{(c'-g')}(\xi)\approx \frac{1}{3}\alpha_E^{(c-g)}(\xi),\quad \beta_M^{(c'-g')}(\xi)\approx \frac{1}{3}\beta_M^{(c-g)}(\xi).
\end{equation}
which is in line with expectations. The results show that long-range chiral corrections make significant contributions to the polarizabilities. Additionally, except for $\Sigma_c^+$ and $\Xi_c^{\prime +}$, whose transition magnetic moments $\mu_{\xi^*\to \xi+\gamma}\approx 0$, the magnetic dipole transitions $\mathcal{B}_6^* \to \mathcal{B}_6 + \gamma$ for the other sextet baryons provide a dominant contribution to the magnetic polarizabilities. This effect is more pronounced for singly bottomed baryons, as detailed in Appendix~\ref{appendix:b_baryon_polarizabilities}.

\begin{table}[htbp]
	\caption{The numerical results of spin-$\frac{1}{2}$ singly charmed baryon electromagnetic polarizabilities (in unit of $10^{-4}~\mathrm{fm}^3$). The values in parentheses represent the uncertainties of the results.}
	\label{tab:polarizabilities_numerical_results}
    \centering
    \setlength{\tabcolsep}{3mm}
    \begin{tabular}{c|ccc|cccc}
    \hline\hline
         & $\alpha_E^{(c-g)}$ & $\alpha_E^{(c^\prime-g^\prime)}$ & $\alpha_E^{\text{Tot.}}$ & $\beta_M^{(c-g)}$ & $\beta_M^{(c^\prime-g^\prime)}$ & $\beta_M^{(b^\prime)}$ & $\beta_M^{\mathrm{Tot.}}$   \\ \hline
        $\Sigma_c^{++}$ & $4.05(114)$ & $1.34(38)$ & $5.39(151)$ & $0.40(12)$ & $0.16(5)$ & $3.21(75)$ & $3.78(77)$ \\ 
        $\Sigma_c^{+}$ & $7.15(200)$ & $2.27(64)$ & $9.42(264)$ & $0.71(20)$ & $0.28(8)$ & $0.02(2)$ & $1.01(28)$ \\ 
        $\Sigma_c^{0}$ & $3.42(96)$ & $1.07(30)$ & $4.48(126)$ & $0.34(10)$ & $0.13(4)$ & $2.33(33)$ & $2.81(36)$ \\ 
        $\Xi_{c}^{\prime+}$ & $2.97(84)$ & $1.08(31)$ & $4.05(114)$ & $0.30(9)$ & $0.12(4)$ & $0.05(5)$ & $0.47(13)$\\ 
        $\Xi_{c}^{\prime0}$ & $2.02(57)$ & $0.67(19)$ & $2.69(76)$ & $0.20(6)$ & $0.08(3)$ & $2.08(19)$ & $2.36(20)$ \\ 
        $\Omega_c^{0}$ & $0.63(18)$ & $0.27(8)$ & $0.90(26)$ & $0.06(2)$ & $0.03(1)$ & $1.84(16)$ & $1.94(17)$ \\ \hline\hline
    \end{tabular}
\end{table}

For comparison, we list the electromagnetic polarizabilities of nucleons~\cite{ParticleDataGroup:2020ssz}:
\begin{equation}
	\begin{array}{ll}
{\alpha}_E^p=11.2 \pm 0.4 \times 10^{-4} \mathrm{fm}^3, & {\beta}_M^p=2.5 \pm 0.4 \times 10^{-4} \mathrm{fm}^3, \\
{\alpha}_E^n=11.8 \pm 1.1 \times 10^{-4} \mathrm{fm}^3, & {\beta}_M^n=3.7 \pm 1.2 \times 10^{-4} \mathrm{fm}^3.
\end{array}
\end{equation}
We observe that the electric polarizabilities of most spin-$\frac{1}{2}$ singly charmed baryons are much smaller than those of nucleons, indicating that they are more ``stiff'' in electric fields. The exception is $\Sigma_c^{+}$, whose electric polarizability is close to that of nucleons. For the magnetic polarizabilities, the situation is different. The nearly degenerate states of $\mathcal{B}_6$ and $\mathcal{B}_6^*$ can lead to large magnetic polarizabilities. The magnetic polarizabilities of most spin-$\frac{1}{2}$ singly charmed baryons are similar to those of nucleons, with the exceptions being $\Sigma_c^+$ and $\Xi_c^{\prime +}$, because their transition magnetic moments $\mu_{\xi^* \to \xi + \gamma} \approx 0$.

\section{Summary}\label{sec:DISCUSSION_AND_CONCULSION}

We calculate the electromagnetic polarizabilities of the spin-$\frac{1}{2}$ singly heavy baryons. The analytical expressions are derived up to $\mathcal{O}(p^3)$. For the lack of experimental data, we have to adopt heavy quark symmetry and the quark model to reduce and estimate our LECs.

For the antitriplet $\mathcal{B}_{\bar{3}}$, we find that parity and angular momentum conservation forbid its chiral fluctuations to $\mathcal{O}(p^3)$. Therefore, in our current calculations, $\mathcal{B}_{\bar{3}}$ behaves like a charged point particle with no electromagnetic polarizability. For the spin-$\frac{1}{2}$ sextet $\mathcal{B}_6$, our results indicate that the long-range chiral corrections make significant contributions to the polarizabilities. Furthermore, the nearly degenerate $\mathcal{B}_6$ and $\mathcal{B}_6^*$ lead to a large contribution to the magnetic polarizabilities. \clabel[OP4]{Although it is challenging to determine $\mathcal{O}(p^4)$ corrections rigorously for heavy baryons, previous studies on nucleon systems have shown that $\mathcal{O}(p^3)$ results already qualitatively align well with experimental data, and $\mathcal{O}(p^4)$ corrections do not alter the qualitative conclusions~\cite{Bernard:1993bg,Bernard:1993ry}. Therefore, it is resonable to expect that the $\mathcal{O}(p^3)$ results presented here provide a reliable and meaningful estimate for the polarizabilities of heavy baryons.}

 Our numerical results can be improved with the new experimental results and the new lattice QCD simulation results in the future. Meanwhile, our analytical expressions can help the chiral extrapolation in lattice QCD simulation.

\section*{ACKNOWLEDGMENTS}
We are grateful to Zi-Yang Lin for the helpful discussions. This project was supported by the National
Natural Science Foundation of China (No. 12475137) and by ERC NuclearTheory (Grant No. 885150). The computational
resources were supported by the high-performance computing
platform of Peking University.

\section*{DATA AVAILABILITY}
The data supporting this study's findings are available within the article.

\begin{appendix}
\section{Loop Integrals}\label{appendix:loop_integrals}
To combine propagator denominators, we introduce integrals over Feynman parameters:
\begin{equation}
	\frac{1}{A_1 A_2 \cdots A_n}=\int_0^1 d x_1 \cdots d x_n \delta\left(\sum x_i-1\right) \frac{(n-1)!}{\left[x_1 A_1+x_2 A_2+\cdots x_n A_n\right]^n}.
\end{equation}
To regularize divergent loop integrals, we use the dimensional regularization scheme and expand them around 4-dimensional spacetime. In this way, one can define many of the loop functions that frequently occur in calculations~\cite{Bernard:1995dp,Scherer:2002tk,Hemmert:1996rw}. Here, we list only those that we need:
\begin{equation}
\begin{aligned}
\frac{1}{i} &\int \frac{d^d \ell}{(2 \pi)^d} \frac{\left\{1, \ell_\mu \ell_\nu, \ell_\mu \ell_\nu \ell_\alpha \ell_\beta\right\}}{(v \cdot \ell-\omega-i \epsilon)\left(M_{\chi}^2-\ell^2-i \epsilon\right)}  \\
&=\left\{J_0\left(\omega, M_{\chi}^2\right)\right.,\quad g_{\mu \nu} J_2\left(\omega, M_{\chi}^2\right)+v_\mu v_\nu J_3\left(\omega, M_{\chi}^2\right),\quad \left.\left(g_{\mu \nu} g_{\alpha \beta}+\text { perm. }\right) J_6\left(\omega, M_{\chi}^2\right)+\ldots\right\}.
\end{aligned}
\end{equation}
All loop-integrals can be expressed via the basis-function $J_0$:
\begin{equation}\label{eq:J0_J2_J6}
\begin{aligned}
	J_0\left(\omega, M_{\chi}^2\right) & =-4 L \omega+\frac{\omega}{8 \pi^2}\left(1-2 \ln \frac{M_\chi}{\mu}\right)-\frac{1}{4 \pi^2} \sqrt{M_{\chi}^2-\omega^2} \arccos \frac{-\omega}{M_\chi}+\mathcal{O}(d-4),\\
J_2\left(\omega, M_{\chi}^2\right) & =\frac{1}{d-1}\left[\left(M_{\chi}^2-\omega^2\right) J_0\left(\omega, M_{\chi}^2\right)-\omega \Delta_{\chi}\right], \\
J_6\left(\omega, M_{\chi}^2\right) & =\frac{1}{d+1}\left[\left(M_{\chi}^2-\omega^2\right) J_2\left(\omega, M_{\chi}^2\right)-\frac{M_{\chi}^2 \omega}{d} \Delta_{\chi}\right].
\end{aligned}
\end{equation}
In Eq. \ref{eq:J0_J2_J6} we have used
\begin{equation}
\begin{aligned}
\Delta_{\chi} & =2 M_{\chi}^2\left(L+\frac{1}{16 \pi^2} \ln \frac{M_\chi}{\mu}\right) +\mathcal{O}(d-4),\\
L & =\frac{\mu^{d-4}}{16 \pi^2}\left[\frac{1}{d-4}+\frac{1}{2}\left(\gamma_E-1-\ln 4 \pi\right)\right],
\end{aligned}
\end{equation}
The $\gamma_E=0.557215$ is Euler constant. The scale $\mu$ is introduced in dimensional regularization.

For the spin-averaged forward Compton scattering amplitude, $J_i(-\omega-\delta)$ and $J_i(\omega-\delta)$ always appear symmetrically. Therefore, for simplicity, we define a new $\mathcal{J}$-function:
\begin{equation}
	\mathcal{J}_i(\omega,\delta,M_{\chi}^2)=J_i(\omega-\delta,M_{\chi}^2)+J_i(-\omega-\delta,M_{\chi}^2)
\end{equation}
With $\mathcal{J}_i^{\prime}$ and $\mathcal{J}_i^{\prime \prime}$ we define the first and second partial derivative with respect to $M_{\chi}^2$,

\begin{equation}
\begin{aligned}
\mathcal{J}_i^{\prime}\left(\omega,\delta ,M_{\chi}^2\right) & =\frac{\partial}{\partial\left(M_{\chi}^2\right)} \mathcal{J}_i\left(\omega, \delta,M_{\chi}^2\right) \\
\mathcal{J}_i^{\prime \prime}\left(\omega, \delta,M_{\chi}^2\right) & =\frac{\partial^2}{\partial\left(M_{\chi}^2\right)^2} \mathcal{J}_i\left(\omega,\delta, M_{\chi}^2\right)
\end{aligned}
\end{equation}

\section{The electromagnetic polarizabilities of singly bottom baryons}~\label{appendix:b_baryon_polarizabilities}

The analytical expression for the electromagnetic polarizabilities of singly bottom baryons is basically the same as that for singly charmed baryons. The only difference is that the parameters should be replaced as follows:
\begin{equation}
	g_1\to g_{1,b},\quad g_{3}\to g_{3,b},\quad \delta_1\to \delta_{1,b},\quad C_{\xi}\to C_{\xi,b}.
\end{equation}
Using the same approach as in Sec.~\ref{sec:NUMERICAL_RESULTS}, we determine the parameters as:
\begin{equation}
	g_{1,b}=0.816 \pm 0.082 \pm 0.029,\quad g_{3,b}=0.707 \pm 0.071 \pm 0.025,\quad \delta_{1,b}=20~\mathrm{MeV}.
\end{equation}
The magnetic moment of the $b$ quark cannot be precisely determined due to the lack of experimental or lattice QCD data. However, since the bottom quark is extremely heavy, its magnetic moment should be very small and will not significantly affect the final results. We estimate it as:
\begin{equation}
	\mu_b=(-0.05\pm 0.05)\mu_N.
\end{equation}
The numerical results for the electromagnetic polarizabilities of singly bottom baryons are listed in Table.~\ref{tab:polarizabilities_numerical_results_b}.

\begin{table}[htbp]
	\caption{The numerical results of spin-$\frac{1}{2}$ singly bottom baryon electromagnetic polarizabilities (in unit of $10^{-4}~\mathrm{fm}^3$). The values in parentheses represent the uncertainties of the results.}
	\label{tab:polarizabilities_numerical_results_b}
    \centering
    \setlength{\tabcolsep}{3mm}
    \begin{tabular}{c|ccc|cccc}
    \hline\hline
         & $\alpha_E^{(c-g)}$ & $\alpha_E^{(c^\prime-g^\prime)}$ & $\alpha_E^{\text{Tot.}}$ & $\beta_M^{(c-g)}$ & $\beta_M^{(c^\prime-g^\prime)}$ & $\beta_M^{(b^\prime)}$ & $\beta_M^{\mathrm{Tot.}}$   \\ \hline
        $\Sigma_b^{+}$ & $3.13(85)$ & $1.36(37)$ & $4.49(121)$ & $0.31(9)$ & $0.14(4)$ & $18.13(365)$ & $18.59(366)$ \\ 
        $\Sigma_b^{0}$ & $5.53(149)$ & $2.36(64)$ & $7.89(212)$ & $0.55(15)$ & $0.25(7)$ & $1.45(54)$ & $2.26(58)$ \\ 
        $\Sigma_b^{-}$ & $2.64(71)$ & $1.12(31)$ & $3.76(102)$ & $0.26(8)$ & $0.12(4)$ & $3.41(102)$ & $3.79(103)$ \\ 
        $\Xi_{b}^{\prime 0}$ & $2.30(62)$ & $1.03(28)$ & $3.33(90)$ & $0.23(7)$ & $0.11(3)$ & $1.86(77)$ & $2.20(78)$\\ 
        $\Xi_{b}^{\prime -}$ & $1.56(43)$ & $0.68(19)$ & $2.24(61)$ & $0.16(5)$ & $0.07(2)$ & $2.85(77)$ & $3.08(77)$ \\ 
        $\Omega_c^{-}$ & $0.49(14)$ & $0.23(7)$ & $0.72(20)$ & $0.05(2)$ & $0.02(1)$ & $2.35(69)$ & $2.42(69)$ \\ \hline\hline
    \end{tabular}
\end{table}

\end{appendix}
\bibliography{references}
\end{document}